\documentclass[aps,prd,reprint,superscriptaddress,nofootinbib,longbibliography]{revtex4-2}

\usepackage[utf8]{inputenc}
\usepackage{amsmath,amssymb,tensor}
\usepackage{hyperref,comment}
\usepackage{graphicx}
\usepackage{bm}
\usepackage{epsfig}

\usepackage{color}
\usepackage{physics}
\usepackage{amsmath}
\usepackage{tikz}
\usepackage{mathdots}
\usepackage{yhmath}
\usepackage{cancel}
\usepackage{color}
\usepackage{siunitx}
\usepackage{array}
\usepackage{multirow}
\usepackage{amssymb}
\usepackage{gensymb}
\usepackage{tabularx}
\usepackage{extarrows}
\usepackage{booktabs}
\usetikzlibrary{fadings}
\usetikzlibrary{patterns}
\usetikzlibrary{shadows.blur}
\usetikzlibrary{shapes}

\begin{document}

\title{IR Black Hole Instabilities Trigger Species-Scale Particle Production}

\author{Luis A. Anchordoqui}
\affiliation{Department of Physics and Astronomy, Lehman College, City University of New York, NY 10468, USA}
\affiliation{Department of Physics, Graduate Center, City University of New York, NY 10016, USA}
\affiliation{Department of Astrophysics, American Museum of Natural History, NY 10024, USA}

\author{Alek Bedroya}
\affiliation{Princeton Gravity Initiative, Princeton University, Princeton, NJ 08544, USA}

\author{Dieter L\"ust}
\affiliation{Max--Planck--Institut f\"ur Physik, Werner--Heisenberg--Institut, 80805 M\"unchen, Germany}
\affiliation{Arnold Sommerfeld Center for Theoretical Physics, Ludwig-Maximilians-Universit\"at M\"unchen, 80333 M\"unchen, Germany}

\author{Houri-Christina Tarazi}
\affiliation{Enrico Fermi Institute \& Kadanoff Center for Theoretical Physics, University of Chicago, Chicago, IL 60637, USA}
\affiliation{Kavli Institute for Cosmological Physics, University of Chicago, Chicago, IL 60637, USA}

\begin{abstract}
\noindent We propose a novel UV-IR mechanism in quantum gravity in which black hole instabilities act as a bridge to the most ultraviolet sector of the theory. Specifically, we argue that when black holes reach a critical temperature associated with a light tower of states, they undergo a phase transition that sources particles near the species scale, a threshold beyond which any effective field theory of gravity fails. Using the existing numerical results, we show that production of such particles through Hawking radiation is always subdominant. We also extend our investigation to a large family of dyonic and dilatonic black holes and show how the conclusion depends on the origin of the gauge symmetry, its dilatonic coupling, and the charge under that gauge symmetry.
\end{abstract}

\maketitle

\section{Introduction}

Incorporating gravity into a quantum field theory introduces two universal features. First, there emerges a massless spin-2 particle, the graviton, which couples universally to all forms of energy and momentum. The self-interaction of the graviton at low energies defines a fundamental mass scale known as the Planck scale $M_{{\rm pl}, d}$. Second, the theory predicts the existence of black holes, which can be understood as massive, point-like objects exhibiting strong gravitational screening.

Black holes are unique in that they manifest both thermodynamic and quantum mechanical behavior that can be probed via infrared (IR) field theory near the event horizon. However, the microstates of black holes are intrinsically ultraviolet (UV) in nature since they are one-particle states with masses above $M_{{\rm pl}, d}$. The fact that one can use the IR description to find the density of UV states via black hole entropy~\cite{Hawking:1974rv}, (verified using BPS microstate counting in string theory~\cite{Strominger:1996sh, Maldacena:1997de}) suggests a profound UV-IR connection in any theory of quantum gravity.

This UV-IR interplay can be made more precise. In all known weakly coupled theories of quantum gravity, there exists an energy scale, significantly below $M_{{\rm pl}, d}$, at which a tower of light, weakly coupled states emerges~\cite{Ooguri:2006in}. Remarkably, this low-energy scale also influences the behavior of black holes with masses much higher than $M_{{\rm pl}, d}$. Specifically, when a black hole’s Hawking temperature approaches the mass scale of the tower, it becomes unstable, signalling a transition to a new phase. This phenomenon has motivated the conjecture proposed in Ref.~\cite{Bedroya:2024uva} that there exists a characteristic \textit{black hole scale}, $\Lambda_{\rm BH}$, below the Planck scale, at which Schwarzschild black holes undergo a phase transition to a more stable configuration. This transition necessarily involves a tower of light states with masses on the order of $\Lambda_{\rm BH}$. 

Two illustrative examples of this phenomenon are well known. In theories with large extra dimensions, a tower of Kaluza-Klein (KK) states appears at low energies, and black holes exhibit a Gregory--Laflamme (GL) instability when the temperature of the black hole is of the order of the KK mass scale~\cite{Gregory:1993vy}. Similarly, in weakly coupled string theories, a tower of string excitations emerges, leading to the Horowitz-Polchinski correspondence point and an associated black hole phase transition~\cite{Horowitz:1997jc}. While Horowitz and Polchinski initially argued that such solutions exist only in spacetimes with $d<7$~\cite{Horowitz:1997jc,Chen:2021dsw} non-compact dimensions, subsequent work constructed explicit solutions in $d=7+\epsilon$ that are under perturbative control for small $\epsilon$ by incorporating higher-derivative corrections \cite{Balthazar:2022hno,Balthazar:2022szl}. These normalizable solutions were later identified as a limiting member of a one-parameter non-normalizable solutions which continue to exist in all dimensions $d>7$~\cite{Bedroya:2024igb}. These works allow for the possibility that string stars exist in all spacetime dimensions $d\leq 10$ which was conjectured to be true in~\cite{} however, this claim is not yet proven. We use \textit{string stars} as a broad term to refer to normalizable winding condensate in any spacetime dimension and we reserve \textit{Horowitz-Polchinski solutions} for spacetime dimensions $d\leq 6$ as they exhibit qualitatively different features from string stars in $d\geq 7$. For more on the Gregory-Laflamme instabilities in the context of string stars see~\cite{Emparan:2024mbp,Chu:2025fko, Bedroya:2024igb}.

In this Letter, we propose a dynamical mechanism linking the UV and IR structure of quantum gravity, wherein black hole instabilities—triggered when the temperature of Schwarzschild black holes reaches $\Lambda_{\rm BH}$—lead to the production of highly UV states at the so-called species scale $\Lambda_s$~\cite{Dvali:2007hz,Dvali:2007wp,Dvali:2009ks,vandeHeisteeg:2022btw,Cribiori:2022nke}. Such a transition can be triggered by the expansion of the internal dimensions on cosmological timescales~\cite{Bedroya:2025fwh}. This species scale represents the threshold beyond which any effective field theory (EFT) description of gravity ceases to be valid. To achieve a separation of scales between $\Lambda_{\rm BH}$ and $\Lambda_s$ we assume the presence of a large extra dimension where by large we mean that the size of the extra dimension is much larger than the length scale $\Lambda_s^{-1}$ associated with the quantum gravity cut-off.

Furthermore, we also study this novel mechanism for charged black holes and show that the addition of charges can have drastic effects depending on the type and amount of the charges. To this end, we extend the calculations of GL instability in the existing literature in two major ways which leads to rich physics: {\it (i)}~we consider a class of dyonic black holes that carry point-like and winding charges in the higher-dimensional theory and {\it (ii)}~we  consider dilatonic couplings which are ubiquitous in string theory. 

In summary, we find that when a Schwarzschild black hole undergoes a GL instability, at the end of that instability, about  $M_{{\rm pl}, d}^{d-2}/\Lambda_s^{d-3}$ of the mass of the black hole converts to particles with $\Lambda_s$ energies. Moreover, we explain the effect of adding charges to the black hole. We consider black holes with charge to mass ratios $q_0$ and $q_1$ respectively under 0-form and 1-form gauge symmetries in the higher-dimensional theory.\footnote{The ratios are normalized to be 1 for extremal single-charge black holes.} It is known that for charged black hole with a single charge, $q_0$ makes the black hole more unstable~\cite{Frolov:2009jr} and $q_1$ makes it more stable~\cite{Miyamoto:2007mh} with respect to the GL instability. For black holes with winding charge, there is a threshold $q_1$ such that for $q_1\geq q_{\rm crit}$, there is no GL instability regardless of the size of the extra dimension or $q_0$. We find that there are two important charge thresholds: $Q_{GL}\propto q_{crit} M$ and $Q_{UV}\sim q_{crit} M_{{\rm pl}, d}^{d-2}/\Lambda_s^{d-3}$, where  $Q_{GL}$ is the charge above which GL instability disappears all together and $Q_{UV}$ is the charge above which the charge of the black hole will end up in an almost extremal black hole with no UV particle production.

We show that these conclusions continue to hold for dyonic black holes with dilatonic gauge couplings given by $g_0\propto \exp(-a_0\phi)$ and $g_1\propto \exp(-a_1\phi)$, where $\phi$ is a canonically normalized scalar field in the $(d+1)$-dimensional theory in Planck units.
For black holes that undergo instability, the endpoint of the instability is a non-uniform string which can be roughly thought of as a combination of a higher dimensional black hole and a lower-dimensional charged black hole with charge to mass ratio of $q_{\rm crit}$. The calculations  of the GL instability for the dyonic and general dilatonic black holes are new and highly technical and are included in an accompanying paper~\cite{GLD}.

\section{Important Energy Scales}

We begin by briefly reviewing three energy scales that are fundamental to any theory of quantum gravity.

The first is the \textit{Planck mass}, which characterizes the self-interactions of the graviton at low energies where the Einstein term dominates. In the absence of fine-tuned cancellations, radiative corrections are expected to generate higher-derivative corrections with coefficients of order unity in Planck units. As a result, when the momentum scale of a gravitational process approaches the Planck mass, the EFT description breaks down, and quantum gravitational effects become significant. However, the Planck scale should be viewed only as an upper bound on the quantum gravity cutoff. In many scenarios, additional sources of higher-derivative corrections—such as extra dimensions or stringy effects—can lower the scale at which the EFT ceases to be valid.

For example, in theories with extra dimensions, higher-derivative terms descend from the higher-dimensional theory and are suppressed by powers of the higher-dimensional Planck scale. Likewise, in weakly coupled string theory, the effective action receives stringy corrections with coefficients of order one in string units, rather than Planck units. In both cases, the cutoff of quantum gravity set by the energy scale controlling the higher-derivative expansion can be parametrically lower than the Planck scale. 

This leads to the second scale, known as the \textit{species scale}, $\Lambda_s$~\cite{Dvali:2007hz,Dvali:2007wp,Dvali:2009ks}. It is defined such that higher-derivative operators of the form $\mathcal{R}^n$ (with $2n$ derivatives) are suppressed by coefficients of order $\mathcal{O}(M_{{\rm pl}, d}^{D-2}/\Lambda_s^{2(n-1)})$~\cite{vandeHeisteeg:2022btw,Cribiori:2022nke}. Importantly, while one may impose a lower cutoff by hand, this is merely a formal artifact and not physically motivated.\footnote{See~\cite{Bedroya:2024uva,Bedroya:2024ubj,Calderon-Infante:2025ldq,Castellano:2025ljk} for the discussion of the subtleties related to the energy-scale dependence of the EFT.} 

The third relevant scale is the \textit{black hole scale}, $\Lambda_{\rm BH}$. As explained above, Schwarzschild black holes undergo a phase transition to a more stable phase at this temperature. GL transition in the presence of a large extra dimension is an example of such a transition~\cite{Gregory:1993vy}. Suppose the extra dimension has a size of $L$. When the black hole radius is much smaller than $L$, the black hole develops the  GL instability in which any initial non-uniformity along the compact dimension grows exponentially.

\section{Endpoint of Gregory--Laflamme}

The endpoint of this instability is not immediately obvious. One early proposal was that the system settles into a non-uniform black string wrapping the extra dimension. However, because such configurations possess lower entropy, they cannot be the true thermodynamic endpoint.

Numerical simulations in Ref.~\cite{Kudoh:2004hs,Lehner:2010pn,Figueras:2022zkg} suggest a different outcome: the unstable black string evolves into a sequence of localized higher-dimensional black holes connected by increasingly thin black string segments in a self-similar pattern. As the evolution progresses, the interconnecting strings become thinner and thinner, eventually requiring a resolution by quantum gravity effects within finite time. As a consistency check, we can verify that the higher-dimensional black holes have higher entropies than lower-dimensional ones.

Envision a spacetime with $d$ non-compact dimensions and $p$ large
extra dimensions that have length scale $L$. Within this set up, a
$d$-dimensional Schwarzschild black hole lifts up to a black $p$-brane that is
uniformly wrapped around the $p$-dimensional compact manifold. Suppose that that when the Schwarzschild radius $r_s$ is comparable to the size of the compact dimensions
(i.e., $r_s \sim L$) the GL instability causes the black $p$-brane to break up
into disjoint ($D$-dimensional) black holes that are localized in the extra dimensions. In the process, the black hole is expected to lose $\mathcal{O}(1)$ fraction of its mass to gravitational radiation, however, if we truly treat the black hole as a closed thermodynamical system by placing it in a box, the gravitational radiation is expected to be reabsorbed and form a higher-dimensional black hole.

\begin{figure}[htb!]
    \centering
\includegraphics[width=\linewidth]{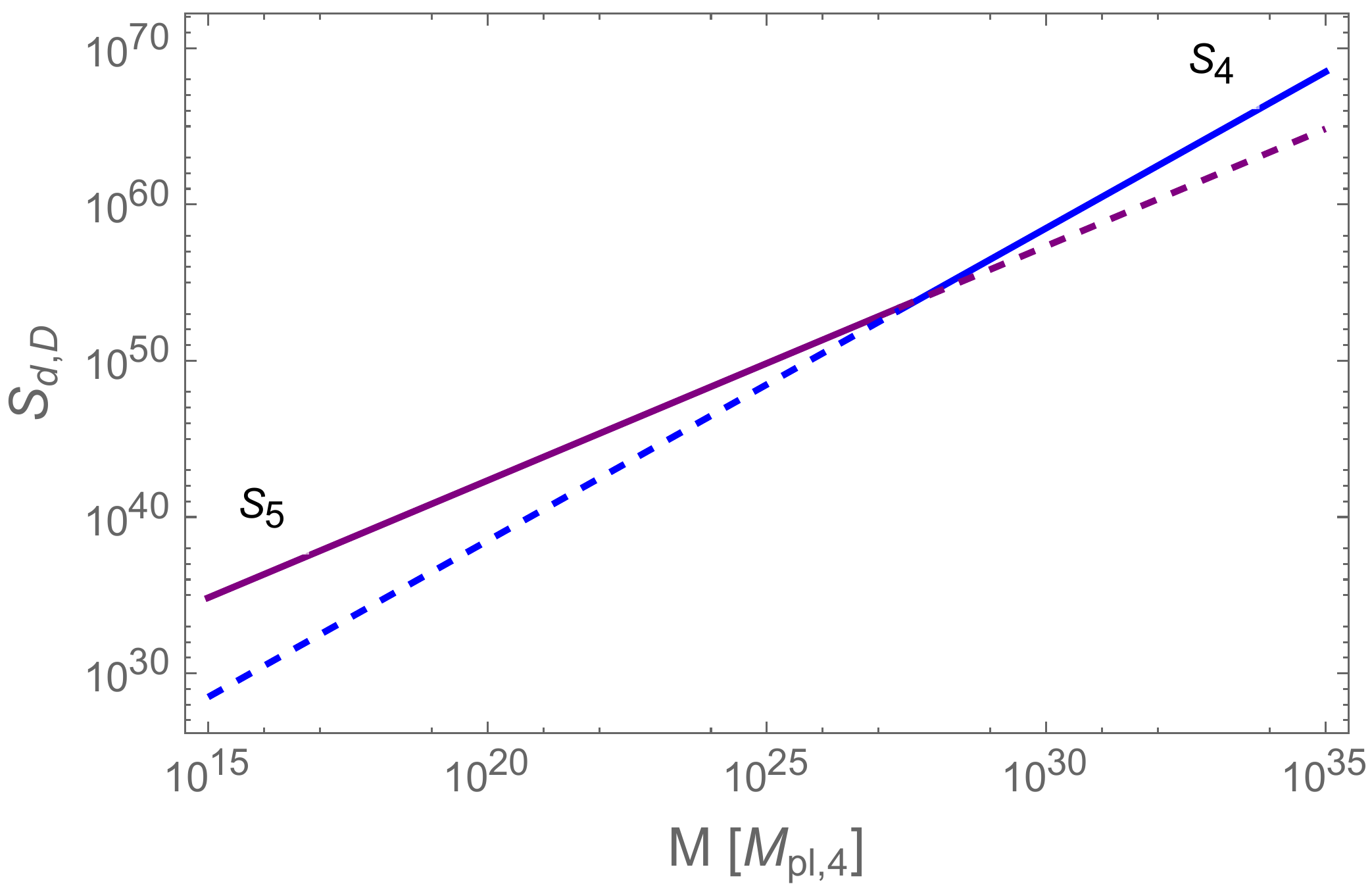}
  \caption{Scaling of the black hole entropy $S_{d,D}$ in $d=4$ and $D=5$ dimensions as a function of the  the black hole mass $M$ in units of the reduced Planck mass $M_{{\rm pl},4}$~\cite{Anchordoqui:2024dxu}. We have taken $L = 1~\mu{\rm m}$ and so  $M_{\Lambda_{\rm BH}} \sim 10^{27}~M_{{\rm pl},4}$.   \label{fig:1}}
\end{figure}

The entropy of the lower-dimensional black hole is found to be
\begin{align}
    S_d= \frac{4\pi^\frac{d+1}{2}}{\Gamma\left(\frac{d-1}{2}\right)}\left(\frac{M}{4\pi M_{{\rm pl}, d}}\right)^\frac{d-2}{d-3}\,,
\end{align}
whereas for the higher-dimensional black hole the entropy is given by
\begin{equation}
    S_D\simeq \frac{4\pi^\frac{D+1}{2}}{\Gamma(\frac{D-1}{2})}\left(\frac{M^{D-2}(2\pi L)^p}{(4\pi)^{D-2} M_{{\rm pl}, d}^{d-2}}\right)^\frac{1}{D-3}\,,
\end{equation}
where $D=p+d$ is the total number of dimensions and $L$ connects the lower-dimensional and higher-dimensional reduced Planck masses
\begin{equation}
    M_{{\rm pl}, d}^{d-2}=M_{{\rm pl}, D}^{D-2}(2\pi L)^p\,,
\label{SD}
\end{equation}
and where, for simplicity, we have assumed that the extra dimensions form a $p$-dimensional torus with radius $L$.
The mass scale at which the lower-dimensional black hole becomes
perturbatively unstable agrees with the mass scale where the entropies
of the two solutions are equal (up to numerical constants): $M\sim L^{d-3}M_{{\rm pl}, d}^{d-2}$. For lower masses, the higher-dimensional black hole is more entropically favorable. In Fig.~\ref{fig:1} we show the GL transition within the context of the dark dimension, a five-dimensional setup
that has a compact space with characteristic length-scale in the micron range (i.e., $d=4$ and $D=5$, with $L= 1\mu{\rm m}$ and $M_{\Lambda_{\rm BH}} \sim 10^{27}~M_{{\rm pl},4}$)~\cite{Montero:2022prj}.  

Now, the entropy of $n$ higher-dimensional black holes, each with a mass of $M/n$, scales as
\begin{equation}
S_{D,n} \sim n \left(\frac{M}{n \ M_{{\rm pl},D}} \right)^{(D-2)/(D-3)} \, .
\end{equation}
Thus, for $d \geq 4$ and $D\geq 5$, we have that $S_{D,n} < S_D$, i.e., the configuration with the highest entropy minimizes the number of higher-dimensional black holes being produced.

If the black hole horizon does not pinch off, it must settle into a stationary black object with a connected horizon. The only viable candidate for such an object is a non-uniform black string. However, the stationary non-uniform black string has lower entropy than the original four-dimensional black hole~\cite{Sorkin:2004qq}. According to the second law of black hole thermodynamics, such a decrease in entropy is forbidden, implying that the horizon must undergo a pinch-off.

\section{Production of $\Lambda_s$ particles}
Classical simulations in general relativity, together with entropy arguments based on the second law of thermodynamics, suggest a common qualitative endpoint of the GL instability: the unstable black string becomes progressively thinner, approaching a naked singularity, while the horizon develops a sequence of localized higher-dimensional black holes that may subsequently merge into a larger black hole. Within classical GR, regarded as an EFT, such an endpoint is not a pathology of the theory but it signals the breakdown of the EFT approximation. In a UV-complete theory, however, the would-be naked singularity should instead be resolved and reinterpreted. More precisely, quantum-gravitational effects are expected to become important once the curvature scale of the thinning black string reaches $\Lambda_s^2$.

This mechanism is particularly transparent in weakly coupled string theory. As the black string thins, the configuration may be approximately described as a set of localized higher-dimensional black holes, corresponding to clumps of energy along the compact direction, together with an increasingly thin nearly uniform black string, which from the lower-dimensional viewpoint is a shrinking black hole. Once its horizon radius reaches the string scale, \(r_h\sim M_s^{-1}\), the system enters the Horowitz--Polchinski transition regime. The black string undergoes a phase transition into a self-gravitating string state, namely a thermal ensemble of highly excited strings. Its mass is parametrically of the same order as that of a Schwarzschild black hole with string-scale radius,
\begin{equation}
M \sim \frac{M_{{\rm pl},d}^{\,d-2}}{M_s^{\,d-3}}.
\end{equation}
This picture is similar in spirit, though not identical, to the analysis of~\cite{Emparan:2024mbp}, who considered the GL instability starting from a configuration already at the Horowitz--Polchinski correspondence point. Their results likewise support the conclusion that the singularity encountered in classical GR is resolved in string theory.

\begin{figure}
    \centering
\includegraphics[width=\linewidth]{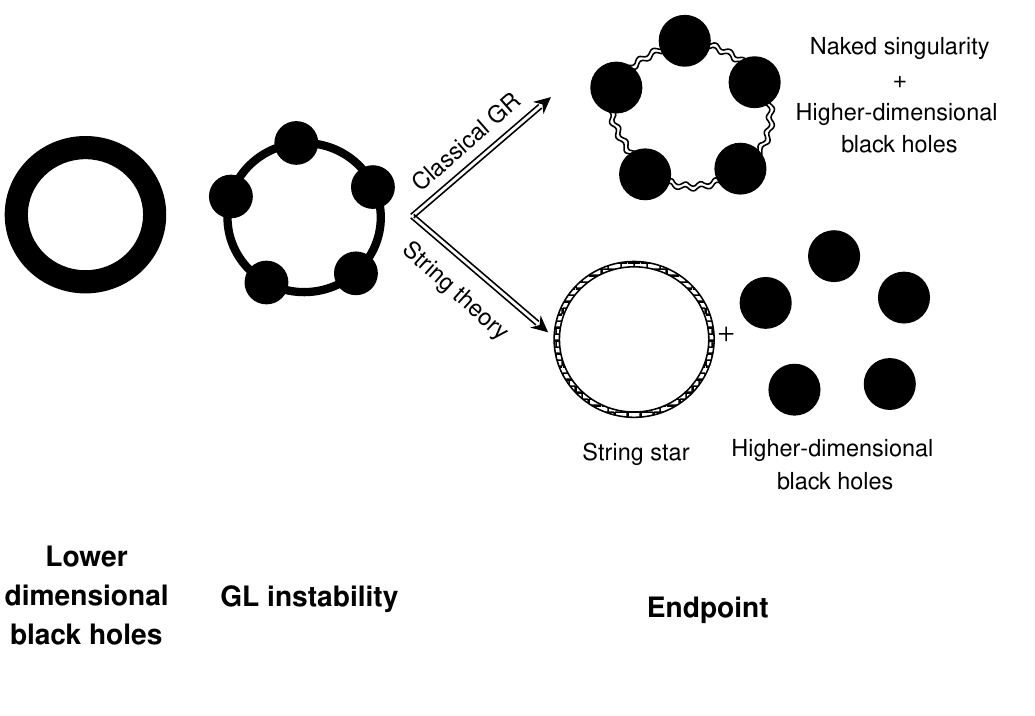}
    \caption{Schematic evolution of the Gregory–Laflamme (GL) instability in classical general relativity and weakly coupled string theory. The circle denotes the compact extra dimension, and the thickness of the shaded region represents the horizon size along it. An initially uniform black string—equivalently a lower-dimensional black hole extended around the circle—develops nonuniformities, forming higher-dimensional black-hole bulges connected by increasingly thin string segments. In classical GR these segments pinch off to a naked singularity, whereas in weakly coupled string theory they are expected to transition into a string star that is uniform along the compact direction.}
    \label{Stringstars}
\end{figure}
The same reasoning should extend beyond weakly coupled string theory. In a general theory of quantum gravity, the string scale is naturally replaced by the species scale \(\Lambda_s\). The thinning black string then continues to shrink until \(r_h\sim \Lambda_s^{-1}\), at which point the gravitational EFT breaks down. Beyond this point, the correct description should be a self-gravitating bound state of quanta with typical energy of order \(\Lambda_s\), which can subsequently decay into stable light particles with the same characteristic energy.

This is the central result of this Letter. Whenever a Schwarzschild black hole undergoes an instability at a temperature \(\Lambda_{\rm BH}\) associated with the lightest tower of states, an amount of order
\begin{equation}
\Delta E \sim \frac{M_{{\rm pl},d}^{\,d-2}}{\Lambda_s^{\,d-3}}
\end{equation}
of the initial black-hole mass is converted into particles with energy \(\Lambda_s\). Remarkably, this quantity is independent of the number of internal dimensions. This provides a striking example of a UV/IR connection: an infrared instability of a black hole can trigger the production of the highest-energy quanta available in the theory, thereby offering, in principle, an observable probe of the deep UV.

Numerical simulations indicate that the GL cascade produces a self-similar hierarchy of localized higher-dimensional black holes connected by ever-thinner string segments~\cite{Lehner:2010pn}. Motivated by this, we model the resulting black-hole mass spectrum by a power law. These black holes may subsequently merge into larger black holes or evaporate individually. It is therefore natural to ask whether evaporation of the low-mass tail of the distribution can provide a larger source of quanta with energy \(\sim \Lambda_s\) than the mechanism proposed above. We now show that this contribution is parametrically subdominant.

Assume that the black holes produced during the cascade are distributed according to a self-similar power law,
\begin{align}
    \rho(m)\equiv m\,n(m)\propto m^{\alpha+1},
\end{align}
where \(n(m)\) is the number density per unit mass and \(\rho(m)\,dm\) is the total mass carried by black holes in the interval \([m,m+dm]\). Requiring the total mass in the distribution to reproduce the mass \(M\) of the initial black hole gives
\begin{align}
    \int_0^M dm\,m\,n(m)\simeq M \
\Rightarrow
    n(m)\simeq (\alpha+2)\frac{m^\alpha}{M^{\alpha+1}},
\end{align}
for \(\alpha>-2\). The total mass stored in black holes lighter than a threshold \(M_c\) is then
\begin{align}
    \Delta M(<M_c)\simeq \int_0^{M_c} dm\,m\,n(m)
    \simeq \frac{M_c^{\alpha+2}}{M^{\alpha+1}}.
\end{align}

We define \(M_c\) as the mass of a five-dimensional black hole whose evaporation time is of the same order as the timescale of the GL cascade. Taking
\begin{align}
    \tau(M)\sim M^{7/2}M_{{\rm pl},5}^{-9/2},
\end{align}
and using \(M_{{\rm pl},5}^3L\simeq M_{{\rm pl},4}^2\), together with the fact that the natural GL timescale is \(\tau_{\rm GL}\sim L\) up to logarithmic corrections, we obtain
\begin{align}
    \tau(M_c)\sim L
    \qquad\Longrightarrow\qquad
    M_c\sim L^{-1/7}M_{{\rm pl},4}^{6/7}.
\end{align}

To estimate the amount of energy converted into quanta of energy \(\Lambda_s\) through Hawking evaporation, we first note that a five-dimensional black hole with radius \(r_h\sim \Lambda_s^{-1}\) has mass
\begin{align}
    m_*\sim M_{{\rm pl},5}^{3}\Lambda_s^{-2}
    \sim M_{{\rm pl},4}^{2}\Lambda_s^{-2}L^{-1}.
\end{align}
This is the maximal amount of energy that any such black hole can release into \(\Lambda_s\)-quanta near the end of its evaporation. A conservative upper bound on the total contribution from the rapidly evaporating tail is therefore
\begin{align}
    \delta M_{{\rm Hawking},s}\lesssim
    m_*\int_{m_*}^{M_c} dm\,n(m).
\end{align}
Evaluating the integral yields
\begin{align}
    \delta M_{{\rm Hawking},s}\sim
    \begin{cases}
\displaystyle
        m_*\frac{M_c^{\alpha+1}}{M^{\alpha+1}},
        & \alpha>-1 ,
    \\[1.2ex]
       \displaystyle
    \frac{m_*^{\alpha+2}}{M^{\alpha+1}},
        & -2<\alpha<-1 .        
        \end{cases}
\end{align}

Using \(M\sim M_{{\rm pl},4}^{2}L\), the energy released by our mechanism,
\begin{align}
    \delta M_{\rm UV}\sim \Lambda_s^{-1}M_{{\rm pl},4}^{2},
\end{align}
dominates over the Hawking contribution. Indeed,
\begin{align}
    \frac{\delta M_{\rm UV}}{\delta M_{{\rm Hawking},s}}
    \sim
    \begin{cases}
     (L\Lambda_s)(LM_{{\rm pl},4})^{\frac{8}{7}(\alpha+1)},
        & \alpha>-1 ,
        \\[1.2ex]
    (L\Lambda_s)^{2\alpha+3},
        & -2<\alpha<-1 .
        \end{cases}
\end{align}
Since \(L\gg \Lambda_s^{-1}\), the ratio is parametrically larger than one provided \(\alpha>-3/2\). Thus, throughout this range, evaporation of the low-mass tail is subdominant, and the dominant source of \(\Lambda_s\)-quanta is the mechanism discussed above.

It is useful to relate $\alpha$ to a simple self-similar model motivated by Ref.~\cite{Lehner:2010pn}. Suppose that, in an idealized bifurcating cascade, each unstable string segment produces one daughter satellite and that the corresponding instability timescales obey $T_{n+1}=c^n \, T_n$, with $0<c<1$. Ref.~\cite{Lehner:2010pn} suggests $c\simeq 1/4$ as a rough late-time estimate. If the instability time is proportional to the local string radius, then the characteristic local length scale satisfies 
\begin{equation}
    r_n\propto T_n\propto c^n,
\end{equation}
and hence the mass scale of the daughter black holes obeys \begin{equation}
    m_n\propto r_n^2\propto c^{2n} . 
\end{equation}

Since the bifurcation produces \(2^n\) black holes at the \(n\)-th stage, the number density per unit mass scales as
\begin{align}
    n(m)\sim m^{\frac{\ln 2}{2\ln c}-1},
\end{align}
and hence
\begin{align}
    \alpha=\frac{\ln 2}{2\ln c}-1.
\end{align}
The condition \(\alpha>-3/2\) is therefore equivalent to
\begin{align}
    c<\frac{1}{2}.
\end{align}

This is comfortably satisfied by the numerical value \(c\simeq 1/4\). We therefore conclude that, for the self-similar GL cascade observed numerically, the dominant production of \(\Lambda_s\)-quanta proceeds through the mechanism proposed above rather than through Hawking evaporation of the low-mass tail. 

\section{Generalization to charged black holes}

The charged case exhibits a qualitatively richer structure. We begin with non-dilatonic charged black holes. In four dimensions, electric-magnetic duality allows us to treat the black hole as electrically charged without loss of generality. However, the corresponding lower-dimensional \(U(1)\) can have two distinct higher-dimensional origins. It may arise either from a charge that can be localized along the compact direction, or from a charge that winds around the compact direction. These two possibilities affect the GL instability in parametrically different ways. Note that in the case of the localized charge, the charge is smeared along the extra dimension to achieve translational symmetry in the extra dimension.

For charges that can be localized, there is no gauge-theoretic obstruction to the pinch-off of the horizon, and sufficiently small nonextremal black holes remain GL unstable for any fixed charge-to-mass ratio \(q_0<1\)~\cite{Bostock:2004mg,Frolov:2009jr}. By contrast, for winding charge the situation is different. If the corresponding charge-to-mass ratio \(q_1\) exceeds a critical value \(q_{\rm crit}\), the GL instability disappears altogether, independently of the size of the black hole~\cite{Miyamoto:2007mh}. There is a simple physical explanation for this difference. In the lower-dimensional theory, the gauge coupling depends on the radion field that controls the size of the extra dimension. For winding charge, the gauge coupling decreases as the size of the extra dimension shrinks, whereas for smeared charge it increases. Since the attractor mechanism drives the near-horizon scalar profile toward smaller effective gauge coupling~\cite{Ferrara:1995ih,Ferrara:1996dd,Ferrara:1996um}, winding charge tends to shrink the extra dimension near the horizon and hence suppress the GL mode, while a smeared local charge drives the radion in the opposite direction and therefore enhances the instability. Moreover, it was shown in Ref.~\cite{Miyamoto:2007mh} that although the exact GL mode must be obtained numerically, the threshold charge-to-mass ratio is easily identified: it occurs precisely when the specific heat changes sign. This is an example of the Gubser--Mitra relation between thermodynamic and classical stability~\cite{Gubser:2000ec,Gubser:2000mm}. At the same time, black holes carrying smeared local charge, which are related by electric-magnetic duality to the winding-charged solutions in the lower-dimensional theory and hence share the same thermodynamic behavior, provide a counterexample to the idea that lower-dimensional thermodynamics alone determines the presence of the GL instability~\cite{Bostock:2004mg}.

The next question is how our analysis of the production of \(\Lambda_s\)-quanta is modified in the presence of charge. For black holes carrying winding charge, the endpoint requires more care. Since the charge threads the compact direction, the string need not pinch off completely. The endpoint is then expected to be a non-uniform string, which can be viewed as a combination of a GL-stable uniform string of sufficiently low mass threading localized higher-dimensional black holes. At first sight, one might then conclude that no \(\Lambda_s\)-quanta are produced, since no naked singularity forms. However, as emphasized above, the appearance of a naked singularity is not the essential point. What matters is the breakdown of the gravitational EFT. Therefore, as long as the curvature scale at the endpoint becomes large enough that the EFT description fails, one should again expect the production of quanta with characteristic energy \(\Lambda_s\).

A simple example is provided by type IIB string theory compactified on a large circle. Consider a black string carrying one unit of wrapped D1-brane charge. The endpoint of the instability is not an ordinary semiclassical black string, but rather a D1-brane configuration. Since such an object does not possess a trustworthy semiclassical horizon, stringy excitations are no longer trapped and can escape. This suggests the following criterion: the production of \(\Lambda_s\)-quanta depends on whether the GL-stable endpoint at charge-to-mass ratio \(q_{\rm crit}\) is still a genuine black object. If the horizon radius of that endpoint is smaller than \(\Lambda_s^{-1}\), then the final state lies beyond the regime of validity of the EFT, and one again expects the emission of an energy of order
\begin{equation}
\Delta E \sim \frac{M_{{\rm pl},d}^{\,d-2}}{\Lambda_s^{\,d-3}}
\end{equation}
in particles of energy \(\Lambda_s\).

Equivalently, if the charge satisfies
\begin{equation}
Q \lesssim Q_{\rm UV} \sim q_{\rm crit}\,\frac{M_{{\rm pl},d}^{\,d-2}}{\Lambda_s^{\,d-3}},
\end{equation}
so that the endpoint is not a macroscopic extremal black brane, and also
\begin{equation}
Q \lesssim Q_{\rm GL} \sim q_{\rm crit}\, M,
\end{equation}
so that the initial black hole is GL unstable, then the instability still leads to the production of \(\Lambda_s\)-quanta.

The single-charge example above is intentionally simple and is not representative of the generic situations encountered in top-down string compactifications. For this reason, in the accompanying paper~\cite{GLD} we also study dyonic black holes carrying both smeared and winding charges in arbitrary spacetime dimension, with gauge couplings of the form
\begin{equation}
g_0 \propto e^{-a_0\phi},
\qquad
g_1 \propto e^{-a_1\phi}.
\end{equation}
Although the full computation of the GL mode in this general setting is technically involved, the results can be summarized succinctly:
\begin{itemize}
    \item The existence of the GL instability depends only on the winding-type charge \(q_1\). 
    \item Increasing the smeared charge to mass ratio $q_0$ decreases the temperature of the smallest stable black hole, but not its radius. 
    \item The critical value \(q_{\rm crit}\) can still be determined analytically from the thermodynamic criterion and is given by
    \begin{equation}
    q_{\rm crit}^{\,d-3}
    =
    \frac{(d-1)a_1^2+d-3}
    {(d-3)(d-2)-(d-1)a_1^2}.
    \end{equation}
    In particular, for sufficiently large $a_1$, even winding charge may cease to stabilize the black hole against the GL instability. Examples and implications in string theory are discussed in Ref.~\cite{GLD}.
    \item Within the unstable regime, increasing the localized charge \(q_0\) makes the black hole more unstable.
\end{itemize}

\begin{figure}
    \centering
    \includegraphics[width=\linewidth]{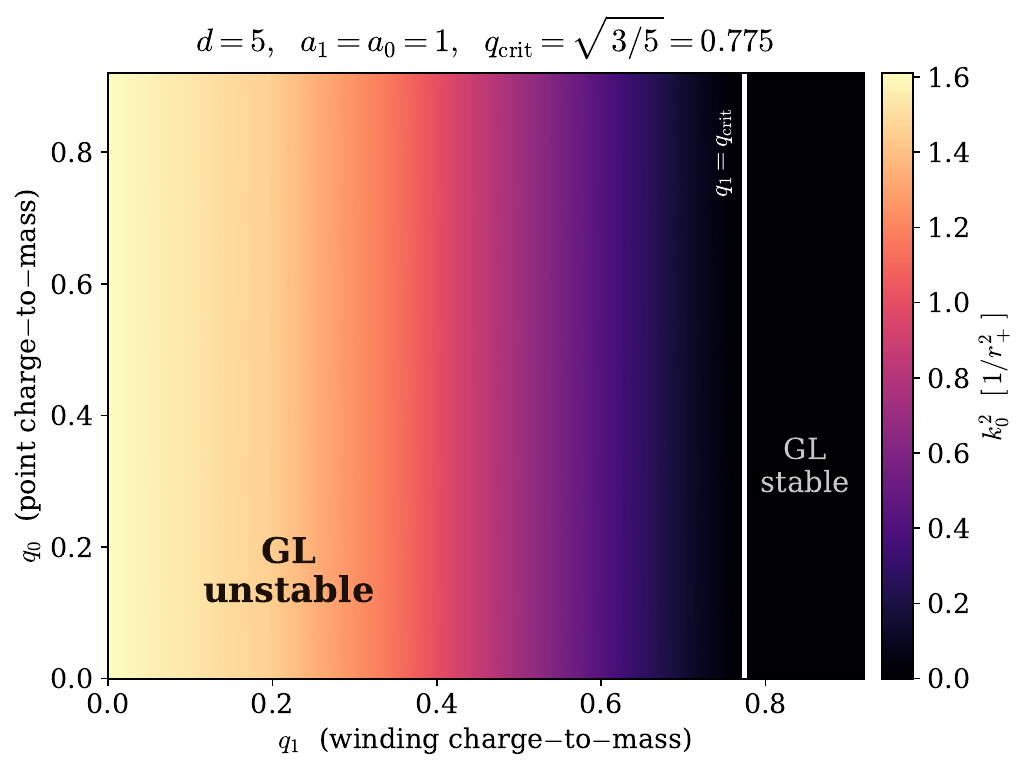}
    \caption{GL-mode eigenvalue $k^2$ as a function of the charge-to-mass ratios $q_1$ and $q_0$ for $d=5$ and $a_0=a_1=1/4$. The white line denotes the threshold $k^2=0$ separating GL-stable and GL-unstable black holes. Since this curve is independent of $q_0$, the existence of the GL instability is controlled solely by the winding charge-to-mass ratio $q_1$. In the unstable region, $k^2>0$, the growth rate nevertheless varies with the smeared charge $q_0$.}
    \label{fig:chargedGL}
\end{figure}
This behavior is illustrated in Fig.~\ref{fig:chargedGL}. There is also a simple way to distinguish winding-type from smeared-type charges. Let \(m(\phi)\) denote the mass of the lightest tower across the moduli space. Then charges satisfying
\begin{equation}
\frac{\nabla g}{g}\cdot \frac{\nabla m}{m} > 0
\end{equation}
and for which the variation of the gauge coupling is not too large admit a finite threshold \(q_{\rm crit}<1\), above which the GL instability disappears. By contrast, for charges satisfying
\begin{equation}
\frac{\nabla g}{g}\cdot\frac{\nabla m}{m}<0
\end{equation}
implies $q_{\rm crit}=1$, so every nonextremal black hole remains Gregory--Laflamme unstable. When the lightest tower is Kaluza--Klein, this criterion provides a natural generalization of the preceding discussion: the sign of the inner product determines whether the compact dimension grows or shrinks near the black-hole horizon and thereby controls the fate of the Gregory--Laflamme instability. Motivated by the relation between light towers and black-hole instabilities found in Ref.~\cite{Bedroya:2024uva}, we conjecture that an analogous criterion also applies when the lightest tower is a string tower. We investigate this conjecture further in Ref.~\cite{GLD}.

We conclude that, for charged black holes, our mechanism for the production of \(\Lambda_s\)-quanta operates whenever the black hole is GL unstable, and the total emitted energy remains parametrically
\begin{equation}
\Delta E \sim \frac{M_{{\rm pl},d}^{\,d-2}}{\Lambda_s^{\,d-3}}.
\end{equation}

\acknowledgments{The work of L.A.A. is supported by the U.S. National Science
Foundation (NSF Grant PHY-2412679). A.B. is supported in part by the Simons Foundation grant number 654561 and by the Princeton Gravity Initiative at Princeton  University. The work of D.L. is supported 
 by the German-Israel-Project (DIP)
on Holography and the Swampland. L.A.A. and A.B. thanks the Harvard Swampland Initiative for hospitality during completion of this work.
 }
\bibliographystyle{utphys.bst}

\bibliography{References2}

\providecommand{\href}[2]{#2}\begingroup\raggedright\begin{thebibliography}{10}

\bibitem{Hawking:1974rv}
S.~W. Hawking, ``{Black hole explosions},'' \href{http://dx.doi.org/10.1038/248030a0}{{\em Nature} {\bfseries 248} (1974) 30--31}.

\bibitem{Strominger:1996sh}
A.~Strominger and C.~Vafa, ``{Microscopic origin of the Bekenstein-Hawking entropy},'' \href{http://dx.doi.org/10.1016/0370-2693(96)00345-0}{{\em Phys. Lett. B} {\bfseries 379} (1996) 99--104}, \href{http://arxiv.org/abs/hep-th/9601029}{{\ttfamily arXiv:hep-th/9601029}}.

\bibitem{Maldacena:1997de}
J.~M. Maldacena, A.~Strominger, and E.~Witten, ``{Black hole entropy in M theory},'' \href{http://dx.doi.org/10.1088/1126-6708/1997/12/002}{{\em JHEP} {\bfseries 12} (1997) 002}, \href{http://arxiv.org/abs/hep-th/9711053}{{\ttfamily arXiv:hep-th/9711053}}.

\bibitem{Ooguri:2006in}
H.~Ooguri and C.~Vafa, ``{On the Geometry of the String Landscape and the Swampland},'' \href{http://dx.doi.org/10.1016/j.nuclphysb.2006.10.033}{{\em Nucl. Phys. B} {\bfseries 766} (2007) 21--33}, \href{http://arxiv.org/abs/hep-th/0605264}{{\ttfamily arXiv:hep-th/0605264}}.

\bibitem{Bedroya:2024uva}
A.~Bedroya, C.~Vafa, and D.~H. Wu, ``{The Tale of Three Scales: the Planck, the Species, and the Black Hole Scales},'' \href{http://arxiv.org/abs/2403.18005}{{\ttfamily arXiv:2403.18005 [hep-th]}}.

\bibitem{Gregory:1993vy}
R.~Gregory and R.~Laflamme, ``{Black strings and p-branes are unstable},'' \href{http://dx.doi.org/10.1103/PhysRevLett.70.2837}{{\em Phys. Rev. Lett.} {\bfseries 70} (1993) 2837--2840}, \href{http://arxiv.org/abs/hep-th/9301052}{{\ttfamily arXiv:hep-th/9301052}}.

\bibitem{Horowitz:1997jc}
G.~T. Horowitz and J.~Polchinski, ``{Selfgravitating fundamental strings},'' \href{http://dx.doi.org/10.1103/PhysRevD.57.2557}{{\em Phys. Rev. D} {\bfseries 57} (1998) 2557--2563}, \href{http://arxiv.org/abs/hep-th/9707170}{{\ttfamily arXiv:hep-th/9707170}}.

\bibitem{Chen:2021dsw}
Y.~Chen, J.~Maldacena, and E.~Witten, ``{On the black hole/string transition},'' \href{http://dx.doi.org/10.1007/JHEP01(2023)103}{{\em JHEP} {\bfseries 01} (2023) 103}, \href{http://arxiv.org/abs/2109.08563}{{\ttfamily arXiv:2109.08563 [hep-th]}}.

\bibitem{Balthazar:2022hno}
B.~Balthazar, J.~Chu, and D.~Kutasov, ``{On small black holes in string theory},'' \href{http://dx.doi.org/10.1007/JHEP03(2024)116}{{\em JHEP} {\bfseries 03} (2024) 116}, \href{http://arxiv.org/abs/2210.12033}{{\ttfamily arXiv:2210.12033 [hep-th]}}.

\bibitem{Balthazar:2022szl}
B.~Balthazar, J.~Chu, and D.~Kutasov, ``{Winding Tachyons and Stringy Black Holes},'' \href{http://arxiv.org/abs/2204.00012}{{\ttfamily arXiv:2204.00012 [hep-th]}}.

\bibitem{Bedroya:2024igb}
A.~Bedroya and D.~Wu, ``{String stars in $d\geq 7$},'' \href{http://arxiv.org/abs/2412.19888}{{\ttfamily arXiv:2412.19888 [hep-th]}}.

\bibitem{Emparan:2024mbp}
R.~Emparan, M.~Sanchez-Garitaonandia, and M.~Toma{\v{s}}evi{\'c}, ``{String theory in a pinch: resolving the Gregory-Laflamme singularity},'' \href{http://dx.doi.org/10.1007/JHEP02(2025)104}{{\em JHEP} {\bfseries 02} (2025) 104}, \href{http://arxiv.org/abs/2411.14998}{{\ttfamily arXiv:2411.14998 [hep-th]}}.

\bibitem{Chu:2025fko}
J.~Chu, ``{Phases of string stars in the presence of a spatial circle},'' \href{http://dx.doi.org/10.1007/JHEP07(2025)080}{{\em JHEP} {\bfseries 07} (2025) 080}, \href{http://arxiv.org/abs/2501.03312}{{\ttfamily arXiv:2501.03312 [hep-th]}}.

\bibitem{Dvali:2007hz}
G.~Dvali, ``{Black Holes and Large N Species Solution to the Hierarchy Problem},'' \href{http://dx.doi.org/10.1002/prop.201000009}{{\em Fortsch. Phys.} {\bfseries 58} (2010) 528--536}, \href{http://arxiv.org/abs/0706.2050}{{\ttfamily arXiv:0706.2050 [hep-th]}}.

\bibitem{Dvali:2007wp}
G.~Dvali and M.~Redi, ``{Black Hole Bound on the Number of Species and Quantum Gravity at LHC},'' \href{http://dx.doi.org/10.1103/PhysRevD.77.045027}{{\em Phys. Rev. D} {\bfseries 77} (2008) 045027}, \href{http://arxiv.org/abs/0710.4344}{{\ttfamily arXiv:0710.4344 [hep-th]}}.

\bibitem{Dvali:2009ks}
G.~Dvali and D.~L{\"u}st, ``{Evaporation of Microscopic Black Holes in String Theory and the Bound on Species},'' \href{http://dx.doi.org/10.1002/prop.201000008}{{\em Fortsch. Phys.} {\bfseries 58} (2010) 505--527}, \href{http://arxiv.org/abs/0912.3167}{{\ttfamily arXiv:0912.3167 [hep-th]}}.

\bibitem{vandeHeisteeg:2022btw}
D.~van~de Heisteeg, C.~Vafa, M.~Wiesner, and D.~H. Wu, ``{Moduli-dependent species scale},'' \href{http://dx.doi.org/10.4310/bpam.2024.v1.n1.a1}{{\em Beijing J. Pure Appl. Math.} {\bfseries 1} no.~1, (2024) 1--41}, \href{http://arxiv.org/abs/2212.06841}{{\ttfamily arXiv:2212.06841 [hep-th]}}.

\bibitem{Cribiori:2022nke}
N.~Cribiori, D.~L{\"u}st, and G.~Staudt, ``{Black hole entropy and moduli-dependent species scale},'' \href{http://dx.doi.org/10.1016/j.physletb.2023.138113}{{\em Phys. Lett. B} {\bfseries 844} (2023) 138113}, \href{http://arxiv.org/abs/2212.10286}{{\ttfamily arXiv:2212.10286 [hep-th]}}.

\bibitem{Bedroya:2025fwh}
A.~Bedroya, G.~Obied, C.~Vafa, and D.~H. Wu, ``{Evolving Dark Sector and the Dark Dimension Scenario},'' \href{http://arxiv.org/abs/2507.03090}{{\ttfamily arXiv:2507.03090 [astro-ph.CO]}}.

\bibitem{Frolov:2009jr}
V.~P. Frolov and A.~A. Shoom, ``{Gregory-Laflamme instability of 5D electrically charged black strings},'' \href{http://dx.doi.org/10.1103/PhysRevD.79.104002}{{\em Phys. Rev. D} {\bfseries 79} (2009) 104002}, \href{http://arxiv.org/abs/0903.2893}{{\ttfamily arXiv:0903.2893 [hep-th]}}.

\bibitem{Miyamoto:2007mh}
U.~Miyamoto, ``{Analytic evidence for the Gubser-Mitra conjecture},'' \href{http://dx.doi.org/10.1016/j.physletb.2007.10.088}{{\em Phys. Lett. B} {\bfseries 659} (2008) 380--384}, \href{http://arxiv.org/abs/0709.1028}{{\ttfamily arXiv:0709.1028 [hep-th]}}.

\bibitem{GLD}
L.~A. Anchordoqui, A.~Bedroya, D.~L\"ust, and H.-C. Tarazi, ``{Gregory--Laflamme for dyonic and dilatonic black holes},'' \href{http://arxiv.org/abs/2603.xxxx}{{\ttfamily 2603.xxxx}}.

\bibitem{Bedroya:2024ubj}
A.~Bedroya, R.~K. Mishra, and M.~Wiesner, ``{Density of states, black holes and the Emergent String Conjecture},'' \href{http://dx.doi.org/10.1007/JHEP01(2025)144}{{\em JHEP} {\bfseries 01} (2025) 144}, \href{http://arxiv.org/abs/2405.00083}{{\ttfamily arXiv:2405.00083 [hep-th]}}.

\bibitem{Calderon-Infante:2025ldq}
J.~Calder{\'o}n-Infante, A.~Castellano, and A.~Herr{\'a}ez, ``{The double EFT expansion in quantum gravity},'' \href{http://dx.doi.org/10.21468/SciPostPhys.19.4.096}{{\em SciPost Phys.} {\bfseries 19} no.~4, (2025) 096}, \href{http://arxiv.org/abs/2501.14880}{{\ttfamily arXiv:2501.14880 [hep-th]}}.

\bibitem{Castellano:2025ljk}
A.~Castellano and M.~Zatti, ``{Black hole entropy, quantum corrections and EFT transitions},'' \href{http://dx.doi.org/10.1007/JHEP08(2025)112}{{\em JHEP} {\bfseries 08} (2025) 112}, \href{http://arxiv.org/abs/2502.02655}{{\ttfamily arXiv:2502.02655 [hep-th]}}.

\bibitem{Kudoh:2004hs}
H.~Kudoh and T.~Wiseman, ``{Connecting black holes and black strings},'' \href{http://dx.doi.org/10.1103/PhysRevLett.94.161102}{{\em Phys. Rev. Lett.} {\bfseries 94} (2005) 161102}, \href{http://arxiv.org/abs/hep-th/0409111}{{\ttfamily arXiv:hep-th/0409111}}.

\bibitem{Lehner:2010pn}
L.~Lehner and F.~Pretorius, ``{Black Strings, Low Viscosity Fluids, and Violation of Cosmic Censorship},'' \href{http://dx.doi.org/10.1103/PhysRevLett.105.101102}{{\em Phys. Rev. Lett.} {\bfseries 105} (2010) 101102}, \href{http://arxiv.org/abs/1006.5960}{{\ttfamily arXiv:1006.5960 [hep-th]}}.

\bibitem{Figueras:2022zkg}
P.~Figueras, T.~Fran{\c{c}}a, C.~Gu, and T.~Andrade, ``{Endpoint of the Gregory-Laflamme instability of black strings revisited},'' \href{http://dx.doi.org/10.1103/PhysRevD.107.044028}{{\em Phys. Rev. D} {\bfseries 107} no.~4, (2023) 044028}, \href{http://arxiv.org/abs/2210.13501}{{\ttfamily arXiv:2210.13501 [hep-th]}}.

\bibitem{Montero:2022prj}
M.~Montero, C.~Vafa, and I.~Valenzuela, ``{The dark dimension and the Swampland},'' \href{http://dx.doi.org/10.1007/JHEP02(2023)022}{{\em JHEP} {\bfseries 02} (2023) 022}, \href{http://arxiv.org/abs/2205.12293}{{\ttfamily arXiv:2205.12293 [hep-th]}}.

\bibitem{Anchordoqui:2024dxu}
L.~A. Anchordoqui, I.~Antoniadis, and D.~L\"ust, ``{More on black holes perceiving the dark dimension},'' \href{http://dx.doi.org/10.1103/PhysRevD.110.015004}{{\em Phys. Rev. D} {\bfseries 110} no.~1, (2024) 015004}, \href{http://arxiv.org/abs/2403.19604}{{\ttfamily arXiv:2403.19604 [hep-th]}}.

\bibitem{Sorkin:2004qq}
E.~Sorkin, ``{A Critical dimension in the black string phase transition},'' \href{http://dx.doi.org/10.1103/PhysRevLett.93.031601}{{\em Phys. Rev. Lett.} {\bfseries 93} (2004) 031601}, \href{http://arxiv.org/abs/hep-th/0402216}{{\ttfamily arXiv:hep-th/0402216}}.

\bibitem{Bostock:2004mg}
P.~Bostock and S.~F. Ross, ``{Smeared branes and the Gubser-Mitra conjecture},'' \href{http://dx.doi.org/10.1103/PhysRevD.70.064014}{{\em Phys. Rev. D} {\bfseries 70} (2004) 064014}, \href{http://arxiv.org/abs/hep-th/0405026}{{\ttfamily arXiv:hep-th/0405026}}.

\bibitem{Ferrara:1995ih}
S.~Ferrara, R.~Kallosh, and A.~Strominger, ``{N=2 extremal black holes},'' \href{http://dx.doi.org/10.1103/PhysRevD.52.R5412}{{\em Phys. Rev. D} {\bfseries 52} (1995) R5412--R5416}, \href{http://arxiv.org/abs/hep-th/9508072}{{\ttfamily arXiv:hep-th/9508072}}.

\bibitem{Ferrara:1996dd}
S.~Ferrara and R.~Kallosh, ``{Supersymmetry and attractors},'' \href{http://dx.doi.org/10.1103/PhysRevD.54.1514}{{\em Phys. Rev. D} {\bfseries 54} (1996) 1514--1524}, \href{http://arxiv.org/abs/hep-th/9602136}{{\ttfamily arXiv:hep-th/9602136}}.

\bibitem{Ferrara:1996um}
S.~Ferrara and R.~Kallosh, ``{Universality of supersymmetric attractors},'' \href{http://dx.doi.org/10.1103/PhysRevD.54.1525}{{\em Phys. Rev. D} {\bfseries 54} (1996) 1525--1534}, \href{http://arxiv.org/abs/hep-th/9603090}{{\ttfamily arXiv:hep-th/9603090}}.

\bibitem{Gubser:2000ec}
S.~S. Gubser and I.~Mitra, ``{Instability of charged black holes in Anti-de Sitter space},'' {\em Clay Math. Proc.} {\bfseries 1} (2002) 221, \href{http://arxiv.org/abs/hep-th/0009126}{{\ttfamily arXiv:hep-th/0009126}}.

\bibitem{Gubser:2000mm}
S.~S. Gubser and I.~Mitra, ``{The Evolution of unstable black holes in anti-de Sitter space},'' \href{http://dx.doi.org/10.1088/1126-6708/2001/08/018}{{\em JHEP} {\bfseries 08} (2001) 018}, \href{http://arxiv.org/abs/hep-th/0011127}{{\ttfamily arXiv:hep-th/0011127}}.

\end{thebibliography}\endgroup

\end{document}